# Extremely low energy ARPES of quantum well states in cubic-GaN/AlN and GaAs/GaAlAs heterostructures

Mahdi Hajlaoui[1], Stefano Ponzoni[1], Michael Deppe[2], Tobias Henksmeier[2], Donat Josef As[2], Dirk Reuter[2], Thomas Zentgraf[2], Claus Michael Schneider[3,4], and Mirko Cinchetti[1]

[1]TU Dortmund University, Department of Physics, Dortmund, 44227, Germany
[2]Department of Physics, Paderborn University, 33098 Paderborn, Germany
[3]Forschungszentrum Jülich, Peter Grünberg Institute (PGI-6), Forschungszentrum Jülich GmbH, Jülich, Germany
[4]Fakultät für Physik and Center for Nanointegration Duisburg-Essen (CENIDE), Germany

## Abstract

Quantum well (QW) heterostructures have been extensively used for the realization of a wide range of optical and electronic devices. Exploiting their potential for further improvement and development requires a fundamental understanding of their electronic structure. So far, the most commonly used experimental techniques for this purpose have been all-optical spectroscopy methods that, however, are generally averaged in momentum space. Additional information can be gained by angle-resolved photoelectron spectroscopy (ARPES), which measures the electronic structure with momentum resolution. Here we report on the use of extremely low energy ARPES (photon energy ~ 7 eV) to increase its depth sensitivity and access buried QW states, located at 3 nm and 6 nm below the surface of cubic-GaN/AlN and GaAs/AlGaAs heterostructures, respectively. We find that the QW states in cubic-GaN/AlN can indeed be observed, but not their energy dispersion because of the high surface roughness. The GaAs/AlGaAs QW states, on the other hand, are buried too deep to be detected by extremely low energy ARPES. Since the sample surface is much flatter, the ARPES spectra of the GaAs/AlGaAs show distinct features in momentum space, which can be reconducted to the band structure of the topmost surface layer of the QW structure. Our results provide important information about the samples' properties required to perform extremely low energy ARPES experiments on electronic states buried in semiconductor heterostructures.

## 1. Introduction:

Semiconductor heterostructures with confined electronic states constitute very intriguing systems in condensed matter physics, both from a point of view of fundamental research[1,2] as well as for device applications[3,4]. Prominent candidates among semiconductor heterostructures are so-called quantum-wells, which can be fabricated by the heteroepitaxial growth of different semiconductor materials. In these quantum wells (QW), a quasi-two-dimensional electron (hole) gas with quantized energy states along the growth direction is formed. The electronic structure of the confined QW states is theoretically described as a series of quasi-2D-parabolic bands[1,2] (subbands) and is usually investigated using all-optical spectroscopy techniques[5] such as absorption, reflectance, luminescence, or magneto-optical spectroscopy. All-optical spectroscopy techniques are in general bulk sensitive, which easily enables the access to the QW states that are usually buried under different layers forming the QW structure. However, the information acquired by all-optical methods is typically averaged in momentum space and a direct visualization of the QW band structure has not been achieved so far. Therefore, it is important to develop a new strategy for extending the experimental investigations to momentum resolved techniques that, at the same time, are sensitive to buried electronic states. This issue is of fundamental interest and can deepen our knowledge of the physics of semiconductor heterostructures, delivering more quantitative and unprecedented insights into their possible future applications.

Angle-resolved photoemission spectroscopy (ARPES) is a very powerful experimental method that gives access to the electronic structure of crystalline materials by detecting the kinetic energy and momentum of the photoemitted electrons [6,7]. It has been successfully applied for investigating surface electronic structures using photon energy in the range of 20-100 eV, where the photoelectron mean free path ($\lambda_{PE}$), which defines the probing depth of this technique, is of the order of 5 Å[8]. The access to more buried subsurface structures requires an increase of $\lambda_{PE}$. For this reason, ARPES has been developed to measure in the soft-X-ray range (300 eV < h$\nu$ < 1500 eV) boosting $\lambda_{PE}$ by a factor of 3-5[9,10]. It has been applied, for example, to study the interface states in many binary oxides[11,12] and strongly correlated materials[13,14] as well as the QW states in hexagonal GaN/AlGaN heterostructures[15] that presents the only ARPES investigation of a semiconductor heterostructure showed so far. However, soft X-ray ARPES suffers from many problems such as the low cross-section and the limited energy and momentum resolution; moreover, it needs to be performed at synchrotron facilities and it has limited application for low energy physics[9]. An alternative approach to increase the depth sensitivity of ARPES, while preserving a high energy and momentum resolution, is the use of photons with very low energy in the deep-UV (DUV) range ($\hbar\omega \cong$ 6-7 eV)[16], also known as ELE (extremely low energy)-ARPES[9, 17-19]. The most important limitation here is that the probed states are made with limited windows of energy (~ 1.5 eV below the Fermi level) and momentum (~ 0.2 Å$^{-1}$ near the center of the first Bouillon zone). Hence, ELE-ARPES is usually applied for specific materials that show interesting physics at the center of the Bouillon zone (BZ) such as the Dirac cone in 3D topological insulators[20,21], kinks in some strongly correlated materials[22,23] and the electronic structure in direct-band gap materials[24]. Interestingly, combining ARPES with two femtosecond laser pulses in pump-probe configuration, where the probe is the DUV energy range[25,26], can provide direct visualization of the photoexcited band structure of both occupied and unoccupied states as well as their ultrafast dynamics in the time domain[20,21,25]. Recently, time-resolved ARPES (trARPES) has been demonstrated as a unique method to measure the dispersion of excitons[27,28] that are more pronounced in QW structures than bulk semiconductors[1]. These peculiarities of ELE-ARPES and their advantages with respect to soft X-ray ARPES and conventional optical spectroscopies motivated us to test and assess its bulk sensitivity for measuring the buried QW states in two semiconductor heterostructures; the cubic-GaN/AlN and GaAs/AlGaAs. The bulk sensitivity will be assessed by comparing ELE-ARPES measurements to standard ARPES experiments performed with photon energy in the range where $\lambda_{PE}$ has its minimum (15 eV-21 eV, the typical photon energy range for ultraviolet photoelectron spectroscopy, UPS). The results can open a new experimental way for achieving detailed knowledge of the QW heterostructures using trELE-ARPES.

## 2. Results
**Growth and QW properties**

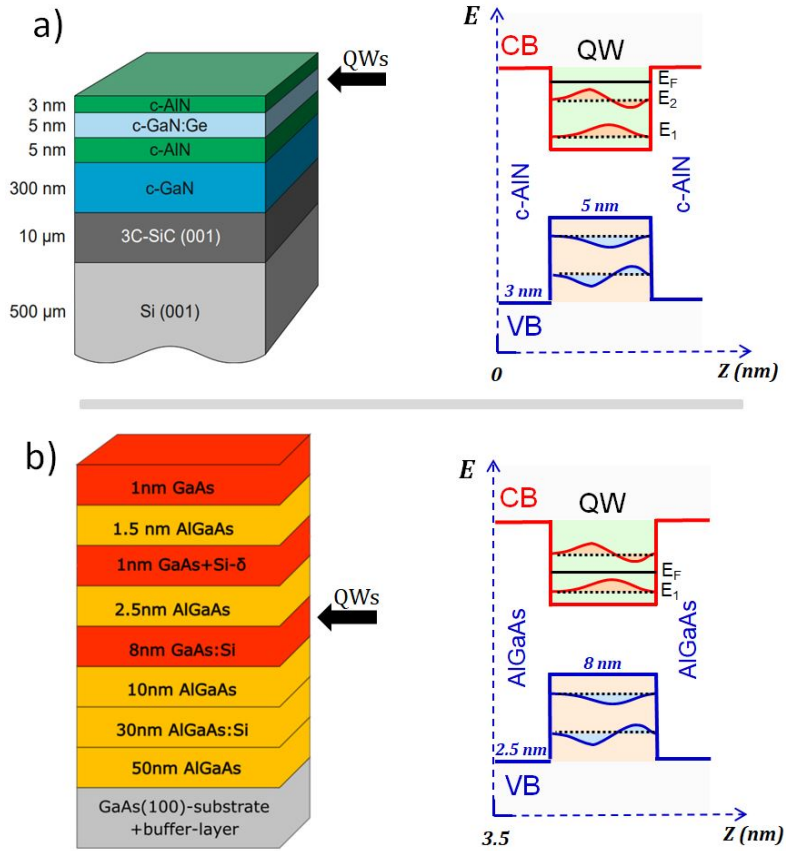

*Figure 1. QW structures fabricated for the ELE-ARPES experiment. a) Schematic overview of the c-GaN/AlN heterostructure (left) and energy band diagram of the single QW including the two occupied CSB states ($E_1$ and $E_2$) located at z=3nm below the sample. b) Same for the GaAs/GaAlAs QW heterostructure, showing one single occupied CSBs ($E_1$) at z=9.5nm below the sample surface.*

We want to assess the depth sensitivity of ARPES using photons with extremely low photon energy in a prototypical system consisting of confined electronic states in single QW heterostructures. As already mentioned, the depth sensitivity of ARPES is determined by $\lambda_{PE}$, a parameter that is strongly material dependent at low photon energy[29], where it is expected to increase upon a decrease in the electron energy. In particular, for the case of semiconductors such as GaAs or GaN, we have estimated the maximal value $\lambda_{PE} > \sim 30$ Å at photon energy hv ~ 6-7 eV[8], which should progressively decrease by increasing the photon energy up to 21 eV.

In order to test this hypothesis, we have fabricated two different QW heterostructures, i.e. c-AlN/GaN[30] and GaAs/AlGaAs, where the QW states are buried at different distances from the sample surface: respectively 3 nm vs. 6 nm. The structure of the QW heterostructures is shown in Fig. 1. Since ARPES measures occupied states, i.e. states below the Fermi level ($E_F$), the QW structures have been designed to ensured n-doping in the QW region, so that at least one conduction subband state (CSbS) is occupied. The c-GaN/AlN and GaAs/AlGaAs structures have been fabricated using respectively plasma-assisted (PA) and solid-source (SS) molecular beam epitaxy (MBE), see Methods. Both samples were grown along the <001> direction of the zinc blend structure, which is the thermodynamic stable phase of the GaAs/AlGaAs system but not for the cubic-GaN/AlN, since GaN and AlN are stable in the wurtzite structure. Figure 1 a) and b) show on the left the structure of the two samples, with the different grown layers (starting from the substrate to the surface) and their thickness. On the right, we show the calculated energy band diagram of the QW states located at 3 nm and 6 nm below the sample surface, respectively, including the calculated Fermi level positions in the conduction band, which ensures the presence of at least one occupied CSbS that we want to detect by ELE-ARPES.

**Surface preparation and characterization**

Even with a relative increase of the bulk sensitivity of ELE-ARPES, a contribution of the surface in the ARPES spectra is always present. Moreover, in this study, comparative ARPES measurements at a very high surface-sensitive condition will be performed as well. This requires an uncontaminated and clean surface during the measurements. With this aim, we protected the surface from oxidation, during the transfer from the MBE to the ARPES setup, by deposing thick amorphous capping layers after growth in the MBE chamber: Arsenic (As) on GaAs/AlGaAs and indium (In) on GaN/AlN. The capping layers were removed by sublimation in the UHV ARPES chamber by annealing at 380°C[31] and 600°C, respectively. The recovery of the surface after annealing was checked by low energy electron diffraction (LEED). Figure 2.a) and 2.d) show respectively the results of LEED of the c-AlN and GaAs surfaces that were taken just prior to the ARPES measurements. The LEED images were taken respectively with a primary electron kinetic energy of 88 eV and 114 eV. In both cases we observe sharp LEED spots, confirming the successful sublimation of the capping layer.

In addition to a clean surface, the roughness of the surface must be minimized for ARPES measurements, because the emission angle, which define the electron momentum, is always measured after the photoelectrons are transmitted through the surface. Therefore, a *flat* surface minimizes photoelectron scattering at the surface and enables the conservation of the electron's momentum parallel to the surface. To address this issue, surface characterization was performed in the MBE chamber by reflection high-energy electron diffraction (RHEED) and *ex-situ* atomic force microscopy (AFM). Figure 2 e) shows a RHEED image taken from the surface of the GaAs/AlGaAs system, where many diffracted spots are following a semicircle arc that indicates a flat single-crystalline surface. However, in the case of c-GaN/AlN system (see Fig. 2 b), the RHEED is observed as "*spotty*" and not elongated streaks, which is a signature of a rougher surface[32]. The same conclusion was obtained, with more quantitative information, using AFM (Fig.2.c and 2.f) that revealed a root mean square (RMS) roughness of ~ 0.39 nm and ~ 2.5 nm on the 10x10 µm$^2$ scale for GaAs and c-AlN surface, respectively.

To summarize, the two studied QW structures have two crucial differences: (i) the distance of the QW states from the sample surface: 3 nm in AlN/GaN vs. 6 nm in GaAs/GaAlAs; (ii) the RMS surface roughness: 2.5 nm in AlN/GaN vs. 0.39 nm in GaAs/AlGaAs on 10x10 µm$^2$. These two parameters are essential to understand the results of the ARPES experiments shown in the following.

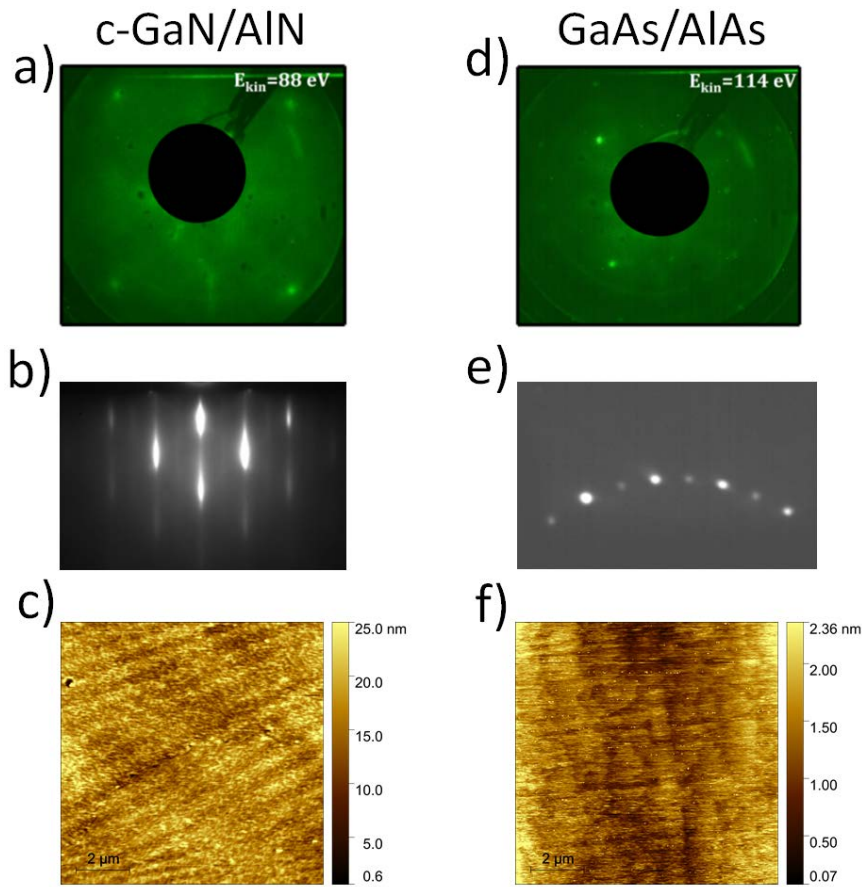

*Figure 2. Characterization of the surface properties of the c-GaN/AlN and GaAs/AlGaAs samples. LEED pattern are shown in a) and d); RHEED in b) and e), and AFM in c) and f).*

**ARPES measurements**

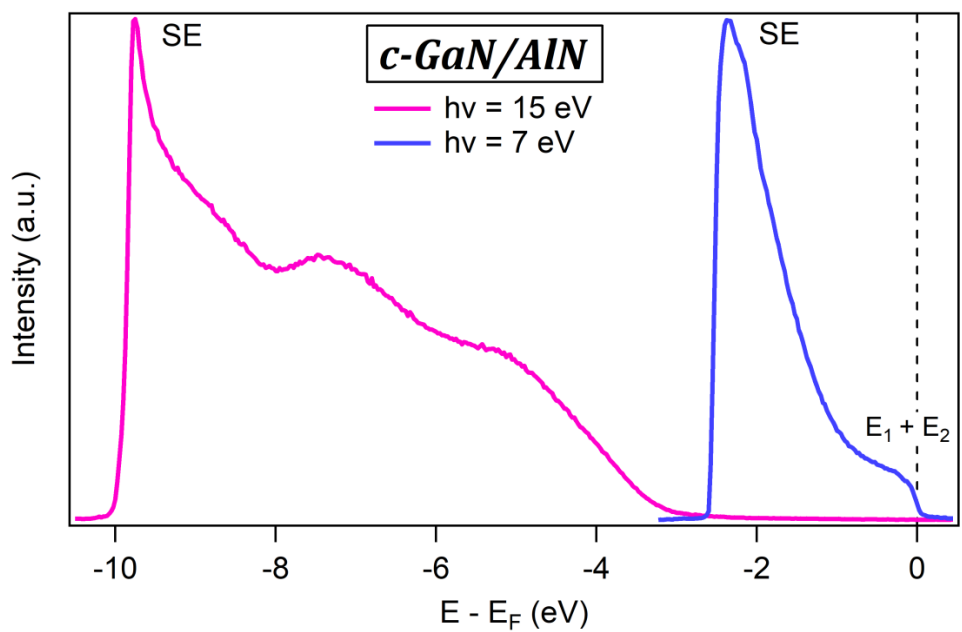

*Figure 3. Angle-integrated ARPES intensity measured from the c-GaN/AlN system with photon energy hv = 7 eV and hv = 15 eV. The energy scale is referenced to the Fermi energy $E_F$.*

We now present the ARPES measurements performed at room temperature at the beamline 5 of the DELTA synchrotron at TU Dortmund University (See methods). Figure 3 shows the angle-integrated photoemission spectra recorded from the c-GaN/AlN samples with photon energy of 15 eV and 7 eV, respectively. As already mentioned, reducing the photon energy to 7eV leads to an increase in bulk sensitivity in the photoemission experiments. Indeed, the 7 eV spectra show a substantial photoemission intensity at the Fermi energy (0 eV in the chosen energy sale) that correspond to the energetic region where the QW states are expected (see Fig.1 a). However, the high roughness of the sample surface introduces a source of elastic and inelastic scattering for the photoemitted electrons, with corresponding loss of energy and momentum information, which prevented us from recording any angle-resolved spectra. The inelastically scattered electrons appear at low kinetic energy in the spectra is the so-called secondary tail (indicated with SE in the figure). The spectrum recorded at 15eV, on the other hand, shows a vanishing photoemission intensity between $E_F$ and -4eV, where the rising edge corresponds to the onset of the valence band of the 3 nm thick AlN surface layer. The additional peak between -6 eV and -8 eV can be ascribed to valence band states of cubic-AlN. As expected, the comparison between the spectra taken at 7 eV and 15 eV allows to confirm the sensitivity of ELE-ARPES to the electronic structure of the QW state buried at 3 nm below the sample surface.

The angle-integrated spectra recorded from the GaAs/AlGaAs sample with photon energy of 7 eV and 21 eV are shown in Fig 4 c. In this case, the recorded photoemission intensity is vanishing close to the Fermi energy in both spectra. As expected, even the ELE-ARPES data recorded with $\hbar\omega \cong 7$ eV cannot detect the QW states, which are indeed buried much deeper below the sample surface than in the c-GaN/AlN sample (6nm vs. 3nm). On the other hand, the much better surface quality of the GaAs/AlGaAs sample allowed us to record very sharp ARPES spectra, as reported in Fig.4.a-b, where it is possible to identify the dispersion of the GaAs valence band. We note that the integrated spectrum at 21 eV is similar to previous PES on GaAs surface[33].

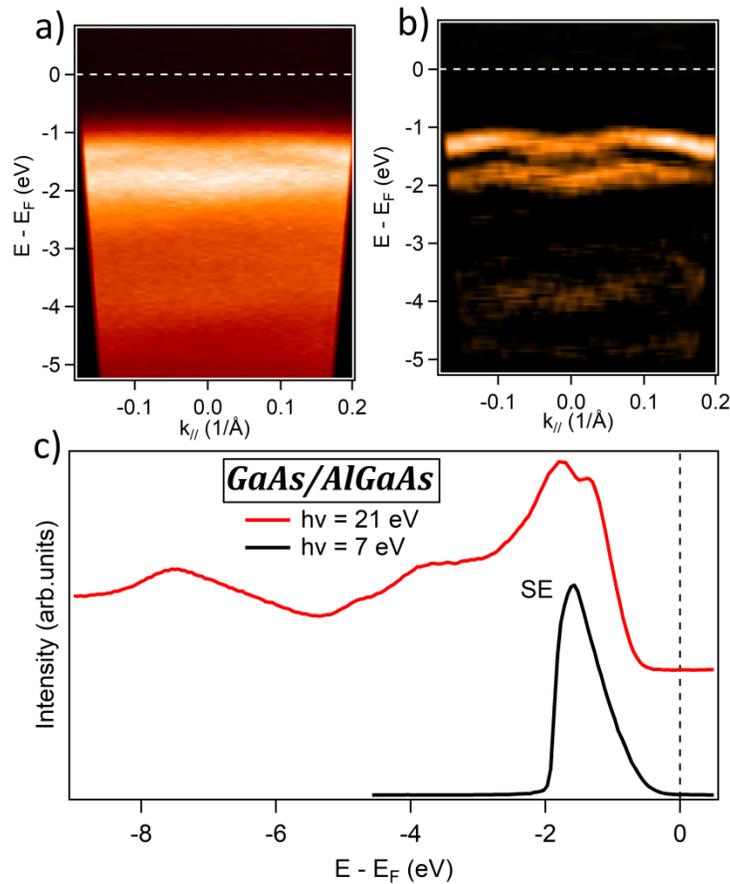

*Figure 4. ARPES measurements of the GaAs/GaAlAs system. a) ARPES spectrum at hv = 21 eV. b) 2$^{nd}$ derivative of the ARPES spectrum in a), displayed for better visualization of the spectral features. c) Angle-integrated ARPES spectra at hv = 7 eV and hv = 21 eV.*

## 3. Discussion

In this paragraph, we discuss the results and clarify the main limitations that prevented the determination of the electronic band structure of the QW states in our ELE-ARPES experiments.

Starting with the c-GaN/AlN heterostructure, we confirmed the enhanced bulk sensitivity of ELE-ARPES (h$\nu$ = 7 eV) that enabled access to the buried QW states located at ~ 3 nm below the sample surface. However, we were not able to determine their angular dispersion in the ARPES spectra because of the high surface roughness. An effort in optimizing the growth conditions to minimize this roughness was made without, however, a significant improvement and the dispersion of the QW states remained undetected. Indeed, the main reason behind this high roughness is the large lattice mismatch (~ -3.2%) between the c-GaN and the 3C-SiC/Si pseudosubstrate, which is the best-used substrate so far for the growth of cubic GaN or AlN[27]. This mismatch introduces dislocations and forms the {111} planar defects or stacking faults (the hexagonal inclusions[34]) that are transferred to the surface resulting in undulations of the growing surface front. A future well-matched substrate better than 3C-SiC could, therefore, flatten the surface of the c-GaN/AlN system, which will permit the observation of the dispersion of the QW states using ELE-ARPES. Contrary to c-GaN/AlN, the surface roughness in GaAs/AlGaAs system was very low as we have demonstrated by AFM and RHEED measurements. This is due to the very low lattice mismatch between the GaAs and AlGaAs materials, and also because the growth of the layers is made in their thermodynamic stable phase. As a result of this flat surface, we have observed the dispersion of the GaAs valence band in the ARPES spectra. However, the QW states remained undetected with ELE-ARPES because they are buried too deep below the heterostructure's surface. Fabricating the GaAs/AlGaAs heterostructure with QW states locate at a lower depth, and at the same time, occupied with electrons is a big challenge. This is a consequence of the presence of a depletion zone close to the sample surface resulting from the conduction band energy pinning in the GaAs system[35]. A further improvement of the growth to minimize the width of the depletion zone might produce the QW at a lower depth, which would enable future ELE-ARPES studies on this system.

## 4. Methods

**Growth of GaN/AlN**

A Riber-32 radio-frequency plasma-assisted molecular beam epitaxy (PAMBE) equipped with standard effusion cells for Ga, In and Al evaporation is used for the growth of the cubic GaN/AlN structures. The nitrogen is delivered by an Oxford plasma source (HD50) and the growth process is in-situ controlled by reflection high energy electron diffraction (RHEED). The growth at a substrate temperature of $T_S$ = 720°C under one monolayer of Ga excess on the surface provides the best sample qualities for cubic GaN (c-GaN) and c-AlN. Deeper insight into the growth of cubic GaN on 3C-SiC/Si (001) can be found in Ref.30. A 100 nm thick c-GaN buffer layer was directly grown on the 3C-SiC substrate. Subsequently, the single QW structure consisting of 5nm cubic AlN bottom barrier, a 5nm cubic GaN well and a 3nm AlN top barrier was evaporated. The top barrier thickness was reduced to 3 nm for the ARPES investigations and the c-GaN QW was homogeneously doped with Ge to a doping level of $7\times10^{19}$cm$^{-3}$. After the growth of the QW structure the sample temperature was reduced to 100°C and covered by a few nm thick In capping layers. This In capping layer was used to protect the QW structure during the transport of the samples from the MBE system to the ARPES system. In the ARPES system the In capping layer was easily reevaporated at a temperature of about 600°C.

**Growth of GaAs/GaAlAs**

The GaAs/AlGaAs heterostructure was grown on a GaAs-(001) substrate by solid-source molecular beam epitaxy (SSMBE). We employed a growth temperature of 590°C and a beam equivalent arsenic (As$_4$) pressure of $p_{As4}$ = $2.25\times10^{-5}$ mbar at a growth rate of 2Ås$^{-1}$ for GaAs, 1Ås$^{-1}$ for AlAs, and 3Ås$^{-1}$ for AlGaAs. First, a buffer layer was grown to provide a smooth initial surface for the heterostructure. The buffer layer consists of a 200 nm thick GaAs layer followed by a 30 periods AlAs (2nm)/GaAs (2nm) superlattice followed by a 50 nm thick GaAs layer. The heterostructure layout (see figure 1) was carefully optimized to obtain states occupied with electrons in a symmetric GaAs quantum well (QW) near the heterostructure surface. The layer sequence after the buffer starts with a 50 nm undoped Al$_{0.3}$Ga$_{0.7}$As layer followed by a 30 nm silicon doped Al$_{0.3}$Ga$_{0.7}$As layer with a doping concentration of around $2.2\times10^{18}$cm$^{-3}$ and another 10nm undoped Al$_{0.3}$Ga$_{0.7}$As layer. Then, 8nm thick silicon doped GaAs layer with a silicon doping concentration of around $3.3\times10^{18}$cm$^{-3}$ is grown; this layer forms the quantum well. It is overgrown by a 2.5nm thick Al$_{0.3}$Ga$_{0.7}$As layer. Then the temperature is lowered to 540°C. A silicon-delta-doping layer with an areal concentration of $2.5\times10^{13}$cm$^{-2}$ is sandwiched between two 0.5nm thick GaAs layers and finally overgrown with 1nm Al$_{0.3}$Ga$_{0.7}$As and 1.5nm GaAs. The lower growth temperature is chosen here to reduce the silicon segregation. After growth of the layer sequence, the temperature is lowered below around 250°C and a thick amorphous arsenic cap layer is deposited to prevent the surface from oxidation outside the chamber.

**Angle resolved photoemission spectroscopy**

The photoemission spectroscopy has been performed at room temperature on the BL 5 at DELTA Synchrotron. It is a VUV beamline supplied by the U250 planar electromagnetic undulator of DELTA. With photon energy tunable between 6 eV and 120 eV, BL 5 is designed for high-resolution photoemission spectroscopy. The experimental station is equipped with a Scienta SES 2002 hemispherical analyzer coupled to a Focus SPLEED detector allowing for both high-resolution angle-resolved photoemission spectroscopy and spin-resolved photoemission spectroscopy.

**Acknowledgments**
We gratefully acknowledge support by the Deutsche Forschungsgemeinschaft in the framework of projects A08, B02, and A06 within the Transregio program, TRR 142 project number 231447078.

**Author contributions**
M.C. and T.Z. conceived the project. M.H performed the ARPES measurements, prepared the figures, and wrote the manuscript. S.P. contributes to the ARPES measurements and the discussion of the ARPES data. M.D. and D.J.A fabricated and characterized the c-GaN/AlN heterostructures. T.H and D.R. fabricated and characterized the c-GaAs/AlGaAs heterostructure. M.C contributed to the writing of the manuscript and the discussion of the ARPES data. All authors reviewed the manuscript.